\documentclass[aps,prb,reprint,groupedaddress,floatfix]{revtex4-1}
\usepackage{graphicx}
\usepackage{epstopdf}
\begin{document}

\preprint{R125 April 17, 2014}

\title{Coupling between octahedral rotations and local polar 
displacements in WO$_3$/ReO$_3$ superlattices}


\author{Joseph T. Schick}
\affiliation{Department of Physics, Villanova University, Villanova, PA 19085}

\author{Lai Jiang}
\author{Diomedes Saldana-Greco}
\author{Andrew M. Rappe}
\affiliation{The Makineni Theoretical Laboratories, Department of Chemistry,
 University of Pennsylvania,
Philadelphia, PA 19104-6323}



\begin{abstract}
We model short-period superlattices of WO$_3$ and ReO$_3$ with
first-principles calculations.  In fully-relaxed superlattices, we
observe that octahedral tilts about an axis in the planes of the 
superlattices do not propagate from one material, despite the 
presence of the corner-shared oxygen atoms.  
However, we find that octahedral rotation is enhanced within
WO$_3$ layers in cases in which strain couples with native
antiferroelectric displacements of tungsten within their octahedral
cages. Resulting structures remain antiferroelectric with low
net global polarization.
Thermodynamic analysis reveals that superlattices with
sufficiently thick ReO$_3$ layers, the absolute number being three or more layers
and the Re fraction $\geq 50\%$, tend to be more stable than the separated
material phases and also show enhanced octahedral rotations in the WO$_3$ layers.
\end{abstract}


\maketitle

\section{\label{section:introduction}Introduction}

Perovskites and related materials
offer a wide variety of electronic, magnetic, and 
structural properties, along with couplings between these
properties that can form the basis for numerous technological
applications.  In most cases, the structures of real perovskites 
deviate from the ideal case, most commonly 
in the form of coherent rotations of the
octahedra in the crystal.\cite{Glazer72p3384}
These rotations break cubic symmetry and typically
require changes in the primitive lattice vectors, as well.
Further symmetry reductions result from other changes in the
structure, including Jahn-Teller and breathing mode distortions 
of the octahedra, and displacements
of the $A$ or $B$ site cations.

Concomitant with the structural deviations are changes in the
electronic, magnetic, and optical properties of the materials.
The size of the semiconducting gap can be
correlated with the degree of octahedral tilt in families of perovskites.
For example, the metal-insulator transition temperature
is correlated with the degree of octahedral
tilt in rare-earth nickelates.\cite{Medarde97p1679}
There is a growing literature on the relationships
between octahedral tilting and properties of interest
including Jahn-Teller distortions and cation 
ordering,\cite{Lufaso04p10} ferroelectricity,\cite{Benedek13p13339}
and multiferroicity in perovskite layered phases and
superlattices.
The emergence of spontaneous polarization in short period
perovskite superlattices due to the combination of two non-polar
rotation patterns, hybrid improper ferroelectricity, has 
been recently discussed as a route to creating novel
multifunctional materials.\cite{Benedek11p107204,Benedek12p11}

Because these useful material properties couple to octahedral tilting, 
there has been an effort to study the
ways in which tilting can be adjusted, by creating
interfaces or layered structures or by applying strain or electric or
magnetic fields.
Octahedral tilting is the primary way perovskites respond to applied
strains because the octahedra are relatively rigid, due to
the covalent $B$-O bonding.\cite{Goodenough55p564,Garcia-Fernandez10p647}
However, the exact structure of any 
perovskite is a result of the interplay between the 
ionic and covalent nature of the bonding
between the $A$- and $B$-cations and the O anions.

In a layered heterostructure,
the tilts and rotations vary with distance from 
the interface, converging to bulk values.
In addition, tilts and rotations will be affected by strains.
The effect of epitaxial strain, in particular, has been previously
investigated in theoretical studies,\cite{Zayak06p094104,Dieguez05p144101} 
including determination of phase diagrams
for numerous perovskites.\cite{Dieguez05p144101}
Strain-induced changes in bond lengths and angles were investigated
experimentally in LaNiO$_3$ on SrTiO$_3$ and LaAlO$_3$, with a 
first principles analysis demonstrating the ability to access
structural phase changes in the response to
tensile strain.\cite{May10p014110}

The dependence of tilt and rotation with respect to distance
from a perturbation 
in La$_{0.75}$Ca$_{0.25}$MnO$_3$ and SrRuO$_3$ 
was recently investigated 
by distorting a single layer of the bulk
material, while relaxing the remaining atoms but fixing the in-plane
lattice parameters to mimic the effect of a 
substrate.\cite{He10p227203} 
These calculations show that La$_{0.75}$Ca$_{0.25}$MnO$_3$ 
sustains out-of-plane tilts for a few layers, and that if the material is 
biaxially compressively strained, the imposed
tilt results in a larger deep-layer tilt 
and a smaller in-plane rotation than found in the same
biaxially, compressively strained material without an imposed tilt layer.
\cite{He10p227203}
However, in SrRuO$_3$ the authors show that imposed tilts have no long-range
propagation into the material, whether or not it is biaxially strained.
\cite{He10p227203}
In both cases, the distance to attain the deep-layer behavior is
very short, of the order of one or two layers.

Certain binary transition-metal oxides form in
a perovskite-like structure, forming $B\mathrm{O}_6$
octhedra with empty $A$-sites.
ReO$_3$ and WO$_3$ possess this structure and allow us to
focus on the $B\mathrm{O}_6$ octahedra exclusively.
WO$_3$ is a monoclinic insulator at room temperature.
WO$_3$ undergoes numerous structural phase transitions both
above and below room temperature, with the triclinic phase stable
below 290~K, for example.\cite{Salje77p574}
The tungsten atoms in WO$_3$ are displaced from their octahedral centers,
in an antiferroelectric pattern.\cite{Rabe13p221}
ReO$_3$ is a conductor with cubic symmetry at room temperature 
and below.\cite{Biltz32p113}
Numerous \emph{ab initio} studies of ReO$_3$ and WO$_3$ have been
carried out,
\cite{Cora97p3945,Wijs99p2684,Walkingshaw04p165110,Wang11p8345,Ping13p165203}
and the link between the WO$_3$ band gap and the polar distortion of
the W atoms has been investigated;\cite{Cora97p3945,Wijs99p2684}
the reduction in energy of the occupied states
is cited as the driver of the polar displacements.
In a recent theoretical investigation of the surfaces of WO$_3$, ReO$_3$,
and surface layers of WO$_3$ on ReO$_3$ and ReO$_3$ on WO$_3$, it
was found that layering is the preferred ordering for Re and
W in bulk WReO$_6$.\cite{Ling11p41}

In this paper, we demonstrate that WO$_3$ and ReO$_3$
both possess longer characteristic decay lengths for out-of-plane
tilts than the reference material CaZrO$_3$, but their in-plane rotation
characteristic decay lengths are comparable.
We also present results for superlattices
and compare their tilt and rotation behaviors to the pure materials.
We determine that the antiferroelectric displacements of tungsten
and the superlattice strain combine to influence the
tilt and rotation angles.

\section{\label{section:computational}Computational details}

Density functional theory (DFT) calculations
were performed with the Perdew-Burke-Ernzerhof (PBE) 
parametrization of the generalized gradient approximation (GGA).
Norm-conserving non-local pseudopotentials were generated with 
the \textsc{opium} code.\cite{Opiumcode}
Calculations were performed with \textsc{quantum espresso}.
\cite{quantumespresso}
The energy cutoff for plane waves was 680~eV.
A $2\times 2\times 2$ Monkhorst-Pack 
$k$-point mesh\cite{Monkhorst1976} was used for relaxing bulk 
materials; a $6\times 6\times 6$ grid was tested to ensure convergence. 
In four-layer supercells a $2\times 2\times 2$ $k$-point grid was used.
In six- and eight-layer supercells, 
a $2\times 2\times 1$ $k$-point sampling was employed.
Ionic forces were converged to $0.0004$~Ry/$a_0$ and pressures in full-cell
relaxations were below 0.2~kbar. Electronic energy was converged 
to $10^{-8}$~Ry. 
Full structural convergence was tested in the 
four-layer superlattices with a $4\times 4\times 4$ $k$-point 
grid.  The bulk ReO$_3$ valence band is composed of strongly delocalized
Re $d$ and O $p$ states.  It has been previously found in LaNiO$_3$, which
is also metallic, that similarly delocalized states strongly screen
the electron-electron interactions.  From this screening it was concluded that 
Hubbard $U$ and hybrid functionals cannot accurately represent the screening 
effects of these delocalized electrons.\cite{Gou11p144101}  For this reason we 
do not include these methods in this calculation.


\section{\label{section:discussion}Results and discussion}

\begin{figure*}
\includegraphics[width=4.61in]{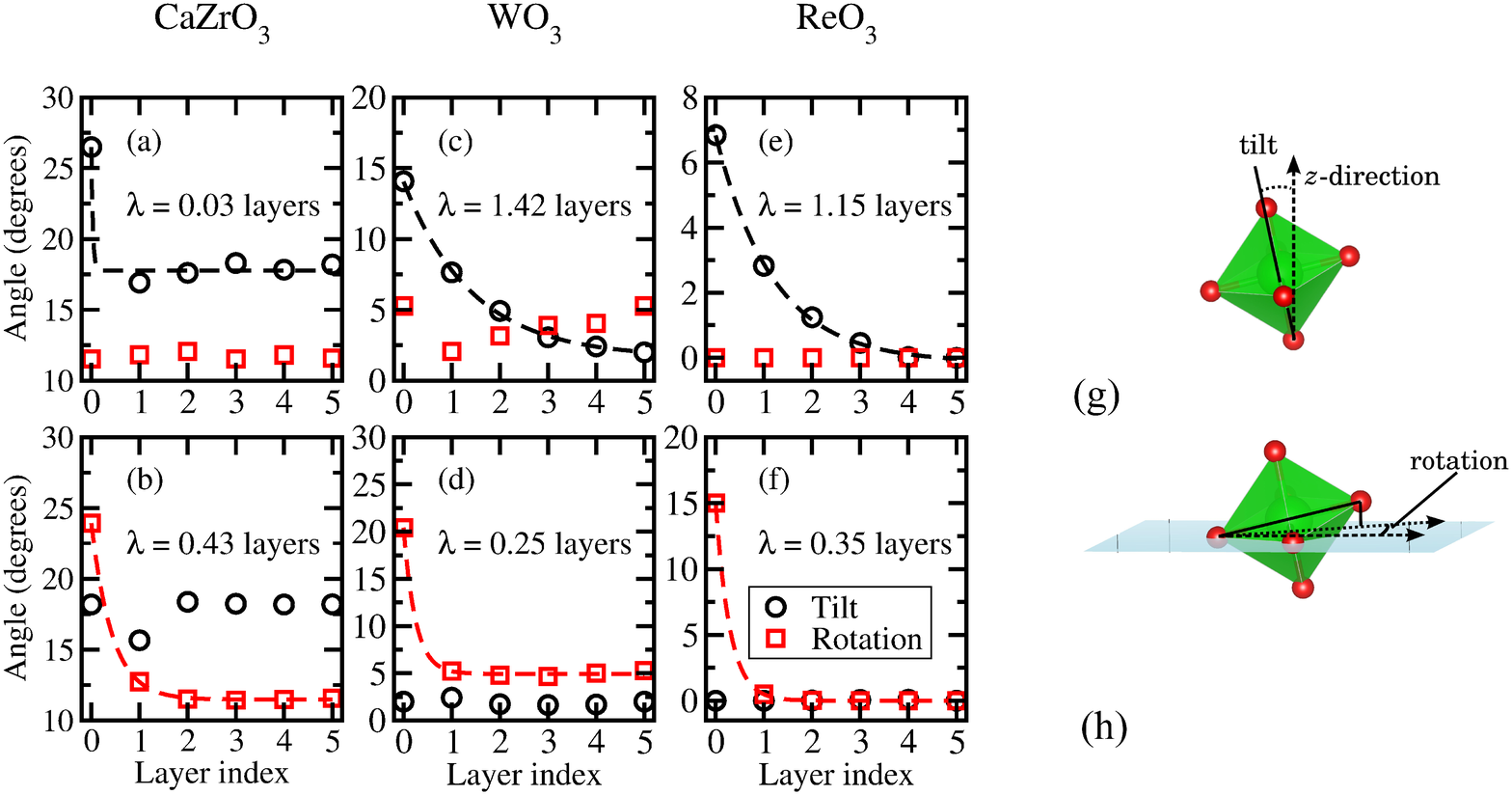}%
\caption{\label{fig:trimat}
(Color online)
Octahedral tilt and rotation angles in each plane along the long axis of
$2\times 2\times 6$ supercells of CaZrO$_3$, WO$_3$, and ReO$_3$. 
The layer zero equatorial oxygen atoms were held fixed to 
maintain an excess octahedral tilt (panels (a), (c), and (e)) or 
octahedral rotation (panels (b), (d), and (f)) alone. 
The equatorial oxygen atoms in layer five were held fixed to maintain 
calculated bulk tilt and rotation angles. 
All other O atoms and all $A$ and $B$ cation locations
were relaxed. The characteristic decay lengths,
$\lambda$, are computed
by fitting to Eq.~\ref{eq:tiltdecay} and given in units of the layer
index.
(g) Tilt angles are the angles between the supercell $c$-axis and 
the polar axis of the octahedra.  (h) Rotation angles evaluated by averaging 
the angles between the supercell $a$- and $b$-axes and $ab$-projected line of 
the nearest octahedral O-O axis.  The differences between the 
lattice directions and the Cartesian axes in this work are small and are 
ignored.
}
\end{figure*}

Imposing tilt on a single layer in CaZrO$_3$
(Fig.~\ref{fig:trimat} panel (a)) has little effect on the material.
In a CaZrO$_3$ $2\times 2\times 6$ unit supercell, we held one layer
fixed at the bulk tilt and rotation values, and tilted or rotated the
oxygen cages on the opposite end. The rest of 
the atoms in the supercell were relaxed.  
The length of decay is characterized with an exponential fit.  
In Fig.~\ref{fig:trimat}-(a) we imposed only a tilt on layer zero and
fit the resulting tilt distribution with the expression
\begin{equation}
 \label{eq:tiltdecay}
 \theta(x) = \theta_\infty + \left(\theta_0 - \theta_\infty \right)
 e^{-x/\lambda}\, ,
\end{equation}
where $\theta$ is the tilt angle, $x$ is the layer index, $\theta_\infty$
is the bulk tilt angle, $\theta_0$ is the imposed tilt at the first layer,
and $\lambda$ is a characteristic length, in units of layers,
for the decay of the tilts and rotations into the bulk. 
In Fig.~\ref{fig:trimat}-(b) the same analysis
is displayed for imposed rotation angle.
Surprisingly, we find that although tilts involve displacement of the
apical oxygen atoms, which are shared by neighboring layers, the
characteristic length $\lambda_\mathrm{tilt}$ in CaZrO$_3$ is very short.
Perhaps equally surprising is the fact that $\lambda_\mathrm{rotate} >
\lambda_\mathrm{tilt}$ in CaZrO$_3$; rotations decay more 
gradually than tilts, though both are quite rapidly decaying.

There is apparently only a limited relationship between the tilt or rotation in
one layer and the next.
However, we observe Fig.~\ref{fig:trimat}-(c) 
through -(f) 
that the characteristic decay lengths for tilt (but not rotation) angles 
in WO$_3$ and ReO$_3$ are significantly longer than for CaZrO$_3$
and that $\lambda_\mathrm{rotate} < \lambda_\mathrm{tilt}$ in the
$B$O$_3$ materials.




\begin{table}
\caption{\label{table:bulklattices}The calculated
parameters of fully relaxed $2\times 2\times 2$ unit cells of ReO$_3$ and 
WO$_3$. (Experimental values are in parentheses.)
Tilt and rotation are defined in Fig.~\ref{fig:trimat}.}
\begin{ruledtabular}
\begin{tabular}{ccc}
Parameter&ReO$_3$&WO$_3$\\
\hline 
$a$&$7.55\, (7.50)\footnotemark[1]$~\AA&$7.46\, (7.31)\footnotemark[2]$~\AA\\
$b$&---&$7.56\, (7.53)\footnotemark[2]$~\AA\\
$c$&---&$7.75\, (7.69)\footnotemark[2]$~\AA\\
$\alpha$&$90^\circ\, (90^\circ)\footnotemark[1]$&$90^\circ\, 
(88.9^\circ)\footnotemark[2]$\\
$\beta$&$90^\circ\, (90^\circ)\footnotemark[1]$&$90.02^\circ\, 
(90.9^\circ)\footnotemark[2]$\\
$\gamma$&$90^\circ\, (90^\circ)\footnotemark[1]$&$90^\circ\, 
(90.9^\circ)\footnotemark[2]$\\
Mean oct.\ tilt&$0^\circ$&$2^\circ$\\
Mean oct.\ rotation&$0^\circ$&$5^\circ$\\
\end{tabular}
\end{ruledtabular}
\footnotetext[1]{From Ref.~\onlinecite{Jorgensen86p4793}.}
\footnotetext[2]{From Ref.~\onlinecite{Woodward95p1305}.}
\end{table}

Having established that tilt and rotation disturbances 
result in layer-to-layer correlations in these binary
materials, we investigate superlattices that are more readily 
synthesized.
Our superlattices are constructed from $2\times 2\times N$
pseudocubic unit cells. Each of the $N$ layers is composed
of either four WO$_6$ or four ReO$_6$ octahedra. The lattice
and all the atomic coordinates are fully relaxed in these calculations.
Calculated lattice parameters of the bulk materials are listed in
Table~\ref{table:bulklattices}.

\begin{figure}
 \includegraphics[width=3.22in]{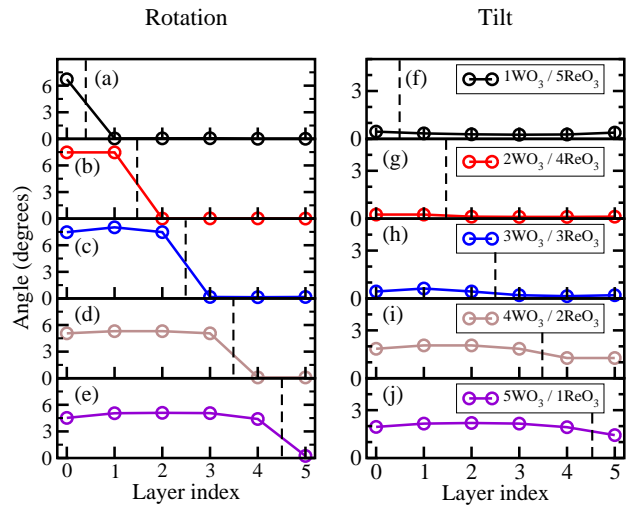}%
\caption{\label{fig:6layersummary}
(Color online)
The rotation (panels a - e) and tilt 
(panels f - j) angles of six-layer
WO$_3$/ReO$_3$ fully-relaxed superlattices are presented as a function of layer
index.  The vertical dashed line indicates the boundary between
WO$_3$ (to the left) and ReO$_3$ (to the right).
ReO$_3$ layers all have nearly zero rotation.
The superlattices with three or fewer WO$_3$ layers
have rotation angles $\approx 8^\circ$.  
The rotation in majority WO$_3$ superlattices are near the bulk values
$\approx 5^\circ$.
All WO$_3$ tilt angles are small, nearly zero in majority 
ReO$_3$ superlattices.  
}
\end{figure}

The rotation and tilt profiles of the six-layer superlattices
are displayed in Fig.~\ref{fig:6layersummary}.
For systems with half or more ReO$_3$ (Fig.~\ref{fig:6layersummary}-(a,b,c)), 
the ReO$_3$ exerts compressive
strain on the WO$_3$ (discussed below), leading to enhanced rotation angles 
$\approx 8^\circ$ in the WO$_3$ layers.
For the two cases with more W than Re, the rotations are near the bulk
value $\approx 5^\circ$.
In all cases, rotations are zero in the
ReO$_3$ layers, with an abrupt change in the octahedral rotation 
angle across the boundary.
The single-layer change in the rotation angle across the superlattice 
boundaries is consistent with the short characteristic length for the decay
of rotations (small values of $\lambda_\mathrm{rotate}$) 
observed previously.

Octahedral tilt angles (Fig.~\ref{fig:6layersummary}-(f - j))
are anticorrelated with rotation angles in WO$_3$;
the systems with larger rotations have lower tilts.
However, unlike the results for rotations, the ReO$_3$
layers also exhibit non-zero tilts for the majority
WO$_3$ superlattices.
The penetration of tilt angles through these superlattices
is consistent with the longer characteristic length
for tilts within both materials
(larger values of $\lambda_\mathrm{tilt}$).

The increase of rotation angles in superlattices with
more ReO$_3$ arise largely from superlattice
strain, with some contribution due to WO$_3$ polar displacements. 
As seen in Table~\ref{table:bulklattices}, two of the WO$_3$ lattice parameters
differ from ReO$_3$ by $1\%$ or $2\%$ with the $b$ lattice 
parameter nearly identical.
The in-plane lattice parameters of the superlattices discussed here
remain close to that of bulk ReO$_3$.  
Because ReO$_3$ has a much larger bulk modulus 
($\approx 200$~GPa)\cite{Pearsall76p1093,Benner77p361} than 
WO$_3$ ($\approx 41$~GPa),\cite{Crichton03p289} the domination of the 
superlattice in-plane lattice parameters by ReO$_3$ is not surprising.
The long dimension in bulk WO$_3$ is a result of minimal tilt $\approx 2^\circ$ 
and the slight elongation of the octahedral axis closest to the direction
of the polar (antiferroelectric) displacement of the W ions. 

The superlattices were initially set up with the $ab$-planes of bulk WO$_3$ 
forming surfaces of contact.
The lattice parameters (Table~\ref{table:bulklattices}) lead us to expect a 
tensile strain on the WO$_3$ layers, 
which would reduce the rotation angle, the opposite of the observed results.
In bulk WO$_3$, calculated W displacements from the geometric centers 
of their octahedral cages are generally toward octahedral edges and make angles 
$\approx 31^\circ$ with respect to the $c$-axis.
However, in the cases with increased rotation (those with more ReO$_3$), 
the directions of the antiferroelectric polar displacements
of the W cations make angles from $\approx 60^\circ$ to $90^\circ$
with respect to the $c$-axis, increasing with the fraction of ReO$_3$.
(See Table~\ref{table:polarcompare}.)
This results in the long axis of the WO$_3$ octahedra also lying in the 
superlattice planes, requiring the WO$_3$ octahedra to rotate more strongly
to fit within the lattice.

\begin{figure}
\includegraphics[width=3.22in]{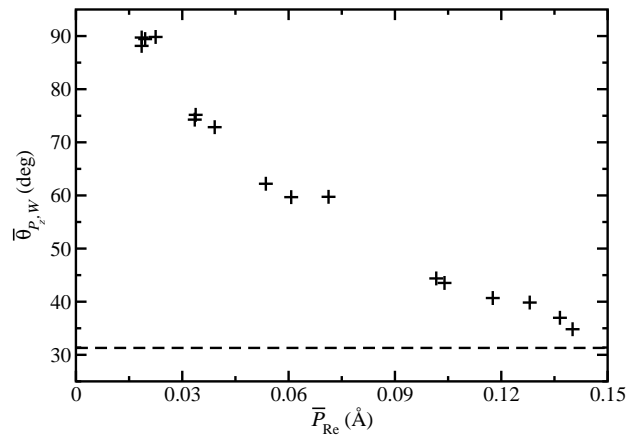}%
\caption{\label{fig:pzvspre} The average angle to the $c$-axis of 
polar displacements in WO$_3$ interface planes vary with the 
average magnitude of antferroelectric polar displacements induced in the 
ReO$_3$ interface planes.  
The horizontal dashed line indicates the average polar displacement angle 
in bulk WO$_3$ ($\approx 31^\circ$).
See Table~\ref{table:polarcompare} for detailed comparison of the data.
}
\end{figure}

The average magnitude of local polar (antiferrelectric)
displacement in the ReO$_3$ layers 
increases as the Re fraction decreases and as the superlattice thickness 
decreases.  Superlattices with a single ReO$_3$ layer have the largest
Re displacements.  The directions of Re displacements are close to
the long axis of the superlattice.
The average magnitude of local polar displacements 
of W atoms are $\approx 0.2$~\AA, 
decreasing to $\approx 0.15$~\AA\, for the highest Re
fractions.  The W polar displacement angle to the superlattice axis 
varies with Re polar displacement at the interface 
plane (Fig.~\ref{fig:pzvspre}), lying completely in the plane for the highest 
Re fraction (Table~\ref{table:polarcompare}).  The polar displacement angles
throughout the WO$_3$ layers differ by $< 1^\circ$ from the surface value. 

To test the effect of the initial conditions on the fully-relaxed
final structures and formation energies of
the superlattices, we also created four- and six-layer superlattices
with the $bc$-planes of bulk WO$_3$ forming the initial contact surfaces.
All possible $m$/$n$ superlattices were formed.  
After full relaxation, the energies of the
test superlattices were higher than the $ab$-oriented superlattices, 
$\approx 0.03$~eV for 
the four-layer and $\approx 0.07$~eV for six-layer superlattices.
The lattice parameters are controlled by the ReO$_3$ material, as discussed
above.
In these tests, rotations do not penetrate into the ReO$_3$ layers 
and the tilt and rotation profiles are consistent with the superlattices 
explored above.  
In these test configurations, when there are odd numbers of WO$_3$ layers,
the initial condition has a net polar shift of W cations along the $b$ 
direction, transverse to the axis of the superlattice.  (The initial polar 
displacement vectors' largest components remain along the $c$-direction.)
These net displacements vanish in the fully-relaxed (WO$_3$)$_5$/(ReO$_3$)$_1$ 
and (WO$_3$)$_3$/(ReO$_3$)$_1$ superlattices but persist in 
the Re-dominated superlattices.  The W-rich superlattices also have
polar displacements closer to the superlattice direction, consistent with
the lower energy superlattices.

\begin{table}
\caption{\label{table:polarcompare}Average antiferroelectric 
displacements of Re in ReO$_3$
surface layers $\overline{P}_\mathrm{Re}$ decrease with increasing ReO$_3$ 
fraction.  
The average angle between the W displacements 
and the axis of the superlattice
in WO$_3$ surface layers
$\overline{\theta}_{P_z,\mathrm{W}}$
generally increases with increasing ReO$_3$ fraction for a given superlattice
thickness.}
\begin{ruledtabular}
\begin{tabular}{cccc}
No.\ of&No.\ of 
&$\overline{P}_\mathrm{Re}$ (\AA)
&$\overline{\theta}_{P_z,\mathrm{W}}$ (deg.)\\
superlattice layers&ReO$_3$ layers &      & \\
\hline
2   &    1     &  0.118  &     41  \\ 
\hline
4   &    1    &   0.128  &     40  \\ 
4   &    2    &   0.071  &     60  \\ 
4   &    3    &   0.022  &     90  \\ 
\hline
6   &    1    &   0.137  &     37  \\ 
6   &    2    &   0.104  &     44  \\ 
6   &    3    &   0.034  &     75  \\ 
6   &    4    &   0.039  &     73  \\ 
6   &    5    &   0.019  &     90  \\ 
\hline
8   &    1    &   0.140  &     35  \\ 
8   &    2    &   0.102  &     44  \\ 
8   &    3    &   0.054  &     62  \\ 
8   &    4    &   0.061  &     60  \\ 
8   &    5    &   0.034  &     74  \\ 
8   &    6    &   0.019  &     88  \\ 
8   &    7    &   0.020  &     89  \\ 
\end{tabular}
\end{ruledtabular}
\end{table}

Above we observed that tilts penetrate across the interfaces 
(Fig.~\ref{fig:6layersummary}), consistent
with the larger $\lambda_\mathrm{tilt}$ in these materials.  As a result,
when ReO$_3$ makes up the majority of the superlattice, tilts are suppressed
within the WO$_3$ layers.  Furthermore, when ReO$_3$ is the majority material,
the induced polar displacements in ReO$_3$ layers are smallest and the polar
displacements in WO$_3$ lie in superlattice layers.  We tested the response of
the polar displacements in bulk WO$_3$ to biaxial strain, of the same amount 
as in the (WO$_3$)$_1$/(ReO$_3$)$_7$ superlattice, and found that the angle of 
the polarization increases only to 
$\approx 40^\circ$ (in comparison to the bulk $\approx 31^\circ$ and the 
superlattice $\approx 89^\circ$). Strain alone
is not enough to cause the large in-plane polar displacement in WO$_3$.

Table~\ref{table:polarcompare} shows that the largest
antiferroelectric Re displacements at the surface occur when there are 
one or two ReO$_3$ layers.  
The superlattices with three or more ReO$_3$ layers have very small surface Re 
displacements and also have W displacements lying closer to the superlattice 
planes, as noted above.
Furthermore, the W displacement angles in all layers in any given superlattice 
remain close to the same value found at the surface.  Evidently, the energy cost 
for displacing the Re cations is large in comparison to the energy cost to 
reorient the W displacements into the superlattice planes.
To estimate this energy, the computed energy required for a 0.14~\AA~ 
antiferroelectric displacement in bulk ReO$_3$, corresponding 
to the maximum displacement
found in the present calculations, is $\approx 0.08$~eV per four formula units.
As will be observed below, this is small in comparison to 
interface and strain energies.

Hybrid improper ferroelectrics have recently been highlighted as a possible 
means to create new multifunctional 
materials.\cite{Benedek12p11,Benedek11p107204}
The mechanism induces a spontaneous polarization through coupling to 
two non-polar rotational modes enabled by creation of $AA'B$O$_3$ 
double-perovskites\cite{Rondinelli12p1918} or Ruddleston-Popper phases.
The induced polarization in hybrid improper ferroelectrics is perpendicular 
to the layering.
We note that the local polarization of the WO$_6$ octahedra lies in or close to 
the layers of Re-majority superlattices (Table~\ref{table:polarcompare}), which 
we have concluded is a result of strain and a microscopic interaction at 
interfaces, a mechanism perhaps related to that discussed for hybrid improper 
ferroelectrics.

\begin{figure}
 \includegraphics[width=3.22in]{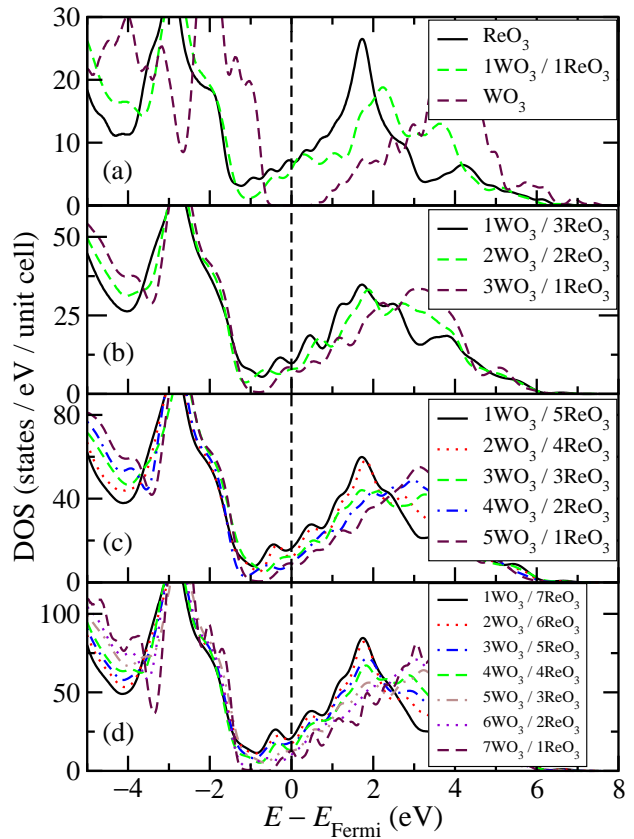}%
\caption{\label{fig:compare-all-dos}
(Color online)
The densities of states for (a) two-layer, (b) four-layer, (c)
six-layer, and (d) eight-layer superlattices 
are compared with the density of states of (a) bulk $B$O$_3$ materials. 
In all calculations there were four formula units in each layer. 
We observe that all the  superlattices possess Fermi levels within bands.  
In the cases with lowest Re fraction, there appears to be a gap opening 
below the Fermi energy.
}
\end{figure}

\begin{figure}
 \includegraphics[width=3.22in]{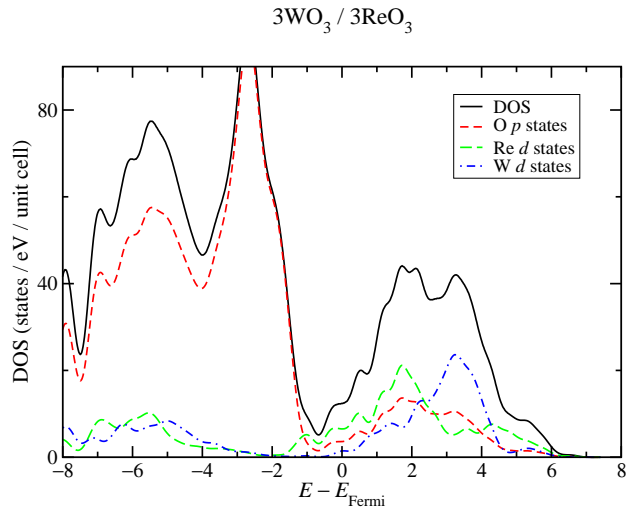}%
\caption{\label{fig:spd-pdos}
(Color online) The densities of states projected onto O-$p$ and metal-$d$
states in the (WO$_3$)$_3$/(ReO$_3$)$_3$ superlattice
shows that the states just around the Fermi energy are made up of
hybridized Re-$d$ and O-$p$ states.
}
\end{figure}

Because there is coupling between the superlattice layers and the
orientation of the W polar displacements, we computed
total polar displacements by summing all the displacement vectors
for each type of $B$-cation in each superlattice.
Total displacements are all of the order $10^{-3}$~\AA\ and less in the low 
energy superlattices.
These displacements are negligible relative to the typical polar
displacements of the W and Re atoms.  
The lack of overall polarization in our 
calculations is perhaps related to the fact that all the superlattices
are conductors. In Fig.~\ref{fig:compare-all-dos} the densities of states 
(DOS) demonstrate the superlattices are metallic, with an apparently
strong contribution from ReO$_3$-like states at the Fermi level.
The DOS projected onto metal-$d$ and O-$p$ states for the 
(WO$_3$)$_3$/(ReO$_3$)$_3$ superlattice (Fig.~\ref{fig:spd-pdos})
shows that hybridized Re-$d$ and O-$p$ states are the strongest
contributions to the DOS at the Fermi energy.
Fig.~\ref{fig:spd-pdos} shows the strong hybridization between O $p$-states 
and Re $d$-states.  Such hybridazation has previously been shown to result
in screening of the electron-electron interaction that cannot be properly
captured with Hubbard $U$ nor with hybrid density 
functionals,\cite{Gou11p144101} justifying our avoidance of
applying such methods in this calculation.

\begin{figure}
\includegraphics[width=3.22in]{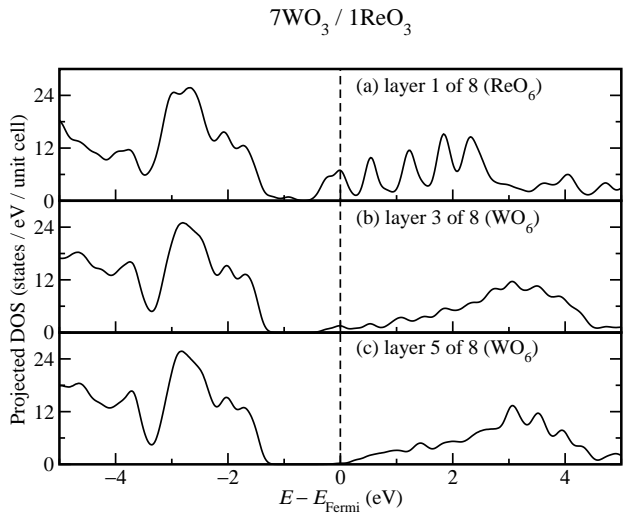}%
\caption{\label{fig:compare-layers-pdos}Projected densities of states for
the (WO$_3$)$_7$/(ReO$_3$)$_1$ superlattice are displayed for three layers
within the superlattice.  In panel (a) is the projection on the ReO$_6$
octahedra, in panel (b) is the projection on the WO$_6$ octahedra of the
second layer after rhenium layer, and in panel (c) is the projection on
the WO$_6$ octahedra on the layer farthest from the rhenium layers.}
\end{figure}

Projecting the densities of states into individual layers of the
superlattice (Fig.~\ref{fig:compare-layers-pdos}) shows that the states 
at the Fermi energy are due primarily to the ReO$_3$ layer of the
superlattice, with the density of states at the Fermi level decreasing rapidly
with distance from the rhenium layer.
We expect reduced conductivity in superlattices with respect to bulk ReO$_3$
because the total density of states at the Fermi energy is lower than in bulk 
ReO$_3$.

\begin{figure}
\includegraphics[width=3.22in]{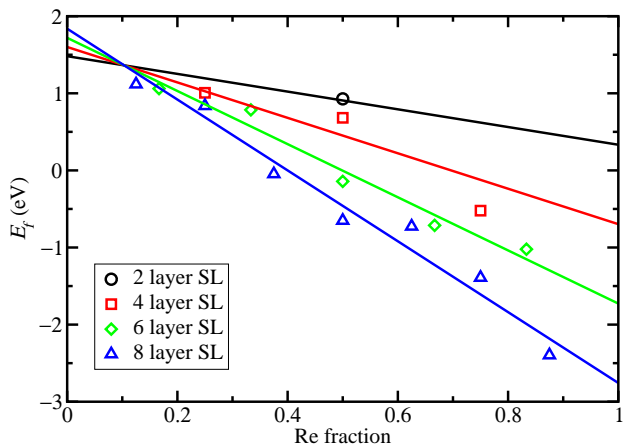}
\caption{\label{fig:compareenergy}
(Color online) The formation energies
$E_f$ relative to separated WO$_3$ and ReO$_3$ phases 
(calculated from Eq.~\ref{eq:eformation}) for all
superlattice calculations are plotted as a function of Re concentration.
Generally superlattices consisting of larger numbers of layers of ReO$_3$
are more stable than those with fewer layers.  The straight lines are energies
calculated from the model (Eq.~\ref{eq:model}) using the values
in Table~\ref{table:fitparms}.
}
\end{figure}

The formation energies of the (WO$_3$)$_m$/(ReO$_3$)$_n$ superlattices 
with respect to equivalent amounts of the bulk materials,
\begin{equation}
\label{eq:eformation}
E_f(m,n) = E_\mathrm{total} - mE_\mathrm{WO_3} - nE_\mathrm{ReO_3}\, ,
\end{equation} 
shows that some superlattices are more stable
than the separated bulk phases. Here $E_\mathrm{WO_3}$ and $E_\mathrm{ReO_3}$
are the formation energies for single layers of the corresponding 
bulk materials.
The data points in Fig.~\ref{fig:compareenergy} display $E_f(m,n)$ for each
superlattice.
We observe that higher rhenium fractions are generally more favorable than
lower, with a preference for three rhenium layers as a minimum.
The possible sources of these energy deviations are 
the energy associated with strains of the materials in the superlattices and
the energy to create the interfaces.
We parameterize the superlattice formation energy in terms 
of the energy costs to strain bulk layers of both WO$_3$ and ReO$_3$ 
($\Delta E_\mathrm{WO_3}^\mathrm{str}$ and $\Delta 
E_\mathrm{ReO_3}^\mathrm{str}$) and 
to create the interfaces $\Delta E_\mathrm{int}$ with the model
\begin{equation}
\label{eq:model}
E_f(m,n) = \Delta E_\mathrm{int} + m \Delta E_\mathrm{WO_3}^\mathrm{str}
+ n \Delta E_\mathrm{ReO_3}^\mathrm{str}\, ,
\end{equation}
where $m$ and $n$ are the numbers of layers of WO$_3$ and ReO$_3$,
respectively.  The results fitting this model are displayed as straight
lines in Fig.~\ref{fig:compareenergy}.  
The fitting parameters in Table~\ref{table:fitparms} show that reducing
the number of layers of WO$_3$ and increasing the number of layers of
ReO$_3$ are both more energetically favorable.  
The majority of the energy change is due to distortion of the ReO$_3$ layers,
as indicated by the magnitude of $\Delta E_\mathrm{ReO_3}^\mathrm{str}$ in 
comparison to $\Delta E_\mathrm{WO_3}^\mathrm{str}$.  The energy
reduction in increasing the Re fraction by changing a layer of WO$_3$ 
to ReO$_3$ (0.574~eV) is also large in comparison 
to the interface energy providing a reason for ReO$_3$ to form thick layers, 
which is consistent with the previous calculation finding that layering is the 
preferred order in WReO$_6$.\cite{Ling11p41}

As noted in the discussion for Table~\ref{table:polarcompare}, increasing 
ReO$_3$ fraction
also decreases the local polarization of ReO$_3$ and increases the angle to the 
superlattice long axis of the local polarization of WO$_3$.
We also tried including polar displacement in the energy model
but found that the fit quality was reduced, suggesting that 
the strain effects are dominant and that
changes in polarization are a response to the constraints imposed by the 
superlattice.
\begin{table}
\caption{\label{table:fitparms}The parameters obtained by fitting 
Eq.~\ref{eq:model} with the energies obtained from DFT results.}
\begin{ruledtabular}
\begin{tabular}{cc}
Parameter&Value (per 4 formula units)\\
\hline 
$\Delta E_\mathrm{int}$&$1.363$~eV\\
$\Delta E_\mathrm{WO_3}^\mathrm{str}$&$0.059$~eV/layer\\
$\Delta E_\mathrm{ReO_3}^\mathrm{str}$&$-0.515$~eV/layer\\
\end{tabular}
\end{ruledtabular}
\end{table}
The energy reduction from increasing ReO$_3$ fraction
is apparently assisted by the ability of the tungsten layers to more readily
accommodate strain through octahedral rotations, which also relates to the 
lower bulk modulus of WO$_3$.
For the superlattices in which ReO$_3$ is the majority material, 
the energies are lower than the bulk phases, suggesting that
these superlattices may be experimentally feasible, either through film
deposition or bulk material segregation.

\section{Conclusion}

In summary, we have found that there is a stronger
layer-to-layer correlation between tilt angles in the binary transition-metal 
oxides WO$_3$ and ReO$_3$, than is found in most studies
of ternary transition-metal oxides.  Rotation angles are enhanced in some
superlattices because antiferroelectric polar distortions 
of the WO$_6$ octahedra cause them
to couple with the layering of the material, reminiscent of the 
hybrid improper ferroelectricity concept.\cite{Dvorak74p1}
However, net global polarization is low.
We find that short-period superlattices are stable relative to bulk 
$X\mathrm{O}_3$ compounds when the ReO$_3$ layer thickness 
is large enough and when ReO$_3$ is the majority species.
This stability suggests that creation of
these short-period superlattices will be experimentally realizable.

\begin{acknowledgments}
The authors gratefully acknowledge their support.
JTS was supported by a sabbatical granted by Villanova University.
LJ was supported by the Air Force Office of Scientific Research under
Grant No. FA9550-10-1-0248.
DSG was supported by the Department of Energy Office of Basic 
Energy Sciences, under grant DE-FG02-07ER15920.
AMR was supported by the Office of Naval Research, under grant 
N00014-12-1-1033.
Computational support was provided by the High Performance 
Computing Modernization Office of the Department of Defense, 
and the National Energy Research Scientific Computing Center of 
the Department of Energy.
\end{acknowledgments}

\end{document}